\documentstyle[preprint,aps]{revtex}
\def\scri{\hbox{${\cal J}$\kern -.645em {\raise .57ex\hbox{$\scriptscriptstyle
(\ $}}}}

\begin{document}
\draft

\title{Topological Censorship}
\author{John L. Friedman}
\address{Institute for Theoretical Physics, University of California, Santa
Barbara, California 93106\\
Department of Physics, University of Wisconsin, Milwaukee,
Wisconsin 53201}
\author{Kristin Schleich and Donald M. Witt}
\address{Department of Physics, University of British Columbia,
Vancouver, British Columbia \ V6T 1Z1}
\date{June 7, 1995}

\maketitle
\begin{abstract}
All three-manifolds are known to occur as Cauchy surfaces of
asymptotically flat vacuum spacetimes and of spacetimes with
positive-energy sources.  We prove here the conjecture that general
relativity does not allow an observer to probe the topology of
spacetime:  any topological structure collapses too quickly to allow
light to traverse it.  More precisely, in a globally hyperbolic,
asymptotically flat spacetime satisfying the null energy condition,
every causal curve from $\scri^-$ to ${\scri}^+$ is homotopic to a
topologically trivial curve from $\scri^-$ to ${\scri}^+$.  (If the
Poincar\'e conjecture is false, the theorem does not prevent one from
probing fake 3-spheres).

\end{abstract}


\narrowtext
Because every three-manifold occurs as the spatial topology of a solution
to the Einstein equations \cite{witt1}, one might ask why such
topological structures are not part of our ordinary experience.  A key
part of the answer is a singularity theorem due to Gannon
\cite{gannon,lee,galloway}, showing that any asymptotically flat
spacetime with a nonsimply connected Cauchy surface has singular time
evolution if it satisfies the weak energy condition.   Only topological
structures comparable in size to the visible universe or small enough
that quantum effects play a crucial role in their dynamics can survive
from the the big bang to the present.

According to the cosmic censorship conjecture \cite{penrose}
singularities forming to the future of a regular initial data surface
are hidden by an event horizon.  If correct, the conjecture suggests
that any topological structures  will ultimately collapse within the
horizon of a set of black holes. This collapse is too rapid to allow
observers to traverse the  wormhole throat for known exact analytically
extended black hole solutions. Consequently one is led to a related
{\it topological} censorship conjecture \cite{friedman} --- that no
observer remaining outside a black hole has time to probe the topology
of spacetime.

A precise formulation of topological censorship requires some standard
definitions related to causal structure \cite{HawkingEllis}.  A
spacetime $M$ is {\it globally hyperbolic} if it has a Cauchy surface
$\Sigma$; that is, $M$ is the domain of dependence $D(\Sigma)$ of a
spacelike hypersurface, $\Sigma$. A spacetime $(M, g_{ab})$ is {\it
asymptotically flat}\cite{gh} if the following conditions hold: (i)
There is a conformal completion $(\widetilde M, \tilde g_{ab})$ where
$\widetilde M$ is compact with $\tilde g_{ab} =\Omega^2 g_{ab}$ for
some
$\Omega$ that vanishes on ${\scri}$ but has null gradient which is
nonvanishing; (ii) The boundary, ${\scri} = {\widetilde M} - M$ is a
disjoint union of past and future parts, ${\scri}^+ \cup {\scri}^-$,
each having topology $S^2\times {\hbox{\rm I\hskip -2.75pt R}}$ with
the ${\hbox{\rm I\hskip -2.75pt R}}$'s complete null
generators. (Topological censorship, however, does not require
completeness in a future direction). In this definition, ${\scri}^+$
and ${\scri}^-$ are future and past null infinity respectively. A {\it
causal curve} is any curve which is non-spacelike. The causal future,
$J^+(S)$, of a set $S$ is the union of $S$ with all points in $M$ that
lie on a future-directed causal curve originating in $S$.  The causal
past, $J^-(S)$, is defined as above with past substituted for future.
The frontier or point set boundary of a set $A \subset X$, with respect
to the set $X$ is given by
${\dot A}= {\overline A} \cap{\overline{X-A}}$ where the bar denotes
the closure of the set in $X$. A useful
property is ${\overline A} ={\rm int}(A)\cup {\dot A}$.
Unless otherwise stated, the frontier will be defined with respect to
$\widetilde M$.

The null energy condition is the requirement that $T_{ab}k^a k^b \geq
0, $ for all null vectors $k^a$.  It is implied by each of the other
common positive  energy conditions: the weak energy condition, the
strong energy condition and the dominant energy
condition\cite{strongE}.  The null energy condition implies that the
convergence of a congruence of null geodesics cannot decrease and thus
that initially converging null geodesics have conjugate points. A
weaker form of energy conditions are averaged energy conditions; their
use for proving the existence of conjugate points is due to Tipler
\cite{tipler}. In particular, Borde proved the {\it averaged} null
energy condition (ANEC) also ensures this property along with even
weaker averages \cite{borde}.

A spacetime satisfies ANEC if the integral of $T_{ab}k^a k^b$ is
nonnegative along every inextendible null geodesic with affine
parameter $\lambda$ and corresponding tangent $k^a$:  $\int d\lambda
T_{ab}k^a k^b \geq 0.$ Finally, we denote by $\gamma_0$ a timelike
curve with past endpoint in ${\scri}^-$ and future endpoint in
${\scri}^+$ that lies in a simply connected neighborhood $U$ of
${\scri}$.

We can now state and prove the topological censorship theorem:

\noindent{\bf Theorem 1}. If an asymptotically flat, globally
hyperbolic spacetime $(M, g_{ab})$ satisfies the averaged null energy
condition, then every causal curve from ${\scri}^-$ to ${\scri }^+$ is
deformable to $\gamma _0\ {\rm rel}\ {\scri}$.

The proof is given below after Lemmas 1 and 2. It is similar to the
argument used by Morris, Thorne and Yurtsever (following a suggestion by Don
Page)\cite{mty} to show that 3-dimensional wormholes are not traversable and
to the proof of Gannon's singularity theorem.

The proof of our theorem is by contradiction. We assume there is a
non-deformable causal curve $\gamma$ from $\scri^-$ to $\scri^+$.
This curve will unwrap in the universal covering space of $M$ to yield
a curve which connects two different asymptotic regions in the
universal cover. One then shows that the unwrapped curve must pass
through a trapped surface.  Finally, one shows that if the spacetime is
globally hyperbolic, curves which pass through trapped surfaces cannot
be observed. However, this contradicts the assumption that $\gamma$
was a causal curve.  Because the existence of such a curve leads to a
contradiction, the topology cannot be actively probed.

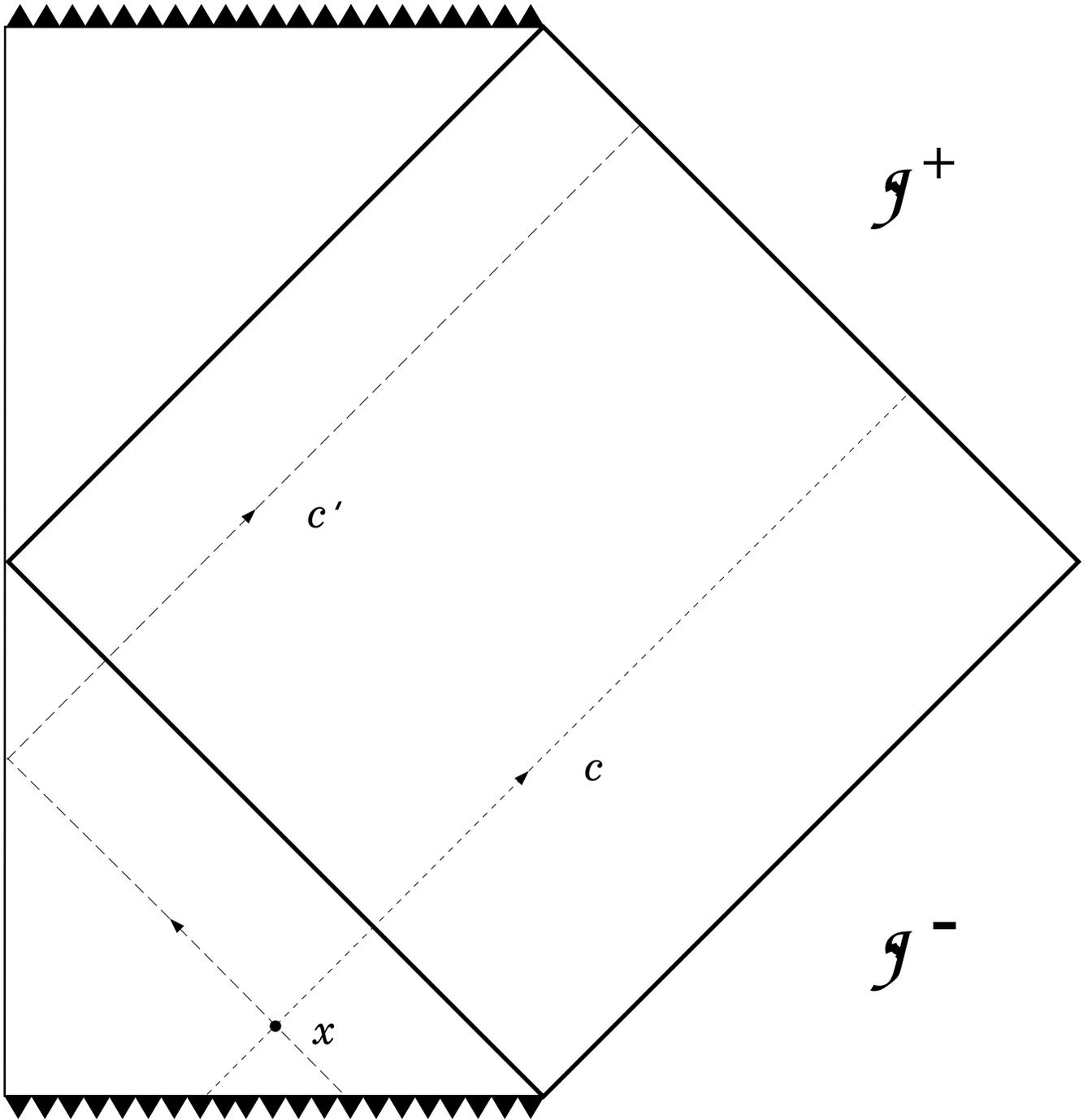
\begin{figure}
\caption{The Penrose diagram for an $RP^3$ geon. Each point in the
diagram is a two-sphere except for the left vertical
boundary, whose points are $RP^2$'s.
\label{pendiag}}
\end{figure}

More precisely, if the Poincar\'e conjecture is true, the only
simply-connected closed 3-manifold is the 3-sphere, and the only
simply-connected, asymptotically flat, globally hyperbolic spacetime
has euclidean topology.  Since the the theorem prevents one from
detecting the first homotopy group of the spacetime, the topological
censorship theorem then prevents one from probing the topology of any
asymptotically flat, globally hyperbolic spacetime.  If the Poincar\'e
conjecture is false, the theorem does not prevent one from probing
fake 3-spheres. (It does, however, prevent one from probing any
topology produced by identifying points on such a fake sphere.)  It may
be helpful to illustrate how the Theorem prevents the detection of
topology by considering the simple example in Fig.~\ref{pendiag} of an
$RP^3$ geon. Choose the $t=0$ slice of Schwarzschild and instead of
extending it across $r=2M$, identify antipodal points at the throat.
The topology of this spatial slice is $RP^3 - pt$. The maximal
evolution of this slice is a spacetime with spatial topology $RP^3 -
pt$. Its universal covering space is the maximally extended
Schwarzschild spacetime. Any non-deformable causal curve unwraps to a
curve which connects the two disconnected asymptotic regions in the
covering space and must pass through a trapped surface.

Since asymptotically flat spacetimes can have multiple disconnected
asymptotic regions, ${\scri }$ in general will have disconnected
components. Let ${\scri }_{\alpha }$ be one such component, and let
$\widetilde {M_{\alpha }}=M \cup {\scri }_{\alpha }$ be a partial
conformal completion, with $\Omega =1$ outside an open neighborhood of
${\scri }_{\alpha }$ which intersects no other component of ${\scri
}$.  The proof uses Prop. 9.2.8 of \cite{HawkingEllis}, slightly
modified as Lemma 2, below, which in turn relies on Lemma 1.

\noindent{\bf Lemma 1}. Let  $({\cal M}, {\rm g}_{ab})$ be any
asymptotically flat spacetime with a simply connected Cauchy surface
$\Sigma$. Let ${\cal T}$ be a smooth closed compact orientable
two-surface in $\Sigma$. Then no null geodesic from ${\cal T}$, inner
directed with respect to ${{\scri }_{\alpha }}$, is part of
${\dot J}^+({\cal T})$.

{\sl Proof of Lemma 1}. Let $t$ be a time function (see
\cite{HawkingEllis} p. 319) for which $\Sigma$ is a surface of constant
$t$, and let ${\cal T}(t)$ be the orbit of ${\cal T}$ under diffeos
generated by $\nabla^a t$.  As $\Sigma$ is simply connected, the
timelike surface ${\cal T}(t)$ separates the spacetime into disjoint
parts, interior and exterior to the surface. Each inner-directed null
geodesic  $\gamma$ from $\cal T$ that meets ${\scri}^+$ must first
intersect ${\cal T}(t)$ for $t > 0$ at a point $p$.  But $p$ is in the
timelike future of $\cal T$: $p\in {I}^+({\cal T})$.  Thus $p$ cannot
lie on ${\dot J}^+({\cal T})$, and $\gamma$ cannot be a generator of
${\dot J}^+({\cal T})$. Q.E.D.

A {\sl congruence} of null geodesics in an open set of a spacetime is a
family of null geodesics such that for each point in the open set there
passes precisely one null geodesic in this family. Of particular
relevance are congruences of null geodesics that are hypersurface
orthogonal to a two-surface $\cal T$. The expansion of such a null
congruence is given by $\theta = s^{ab} \nabla_a k_b$ where $s_{ab}$ is
the metric of the two-surface and $k^a$ is the tangent vector field to
the congruence.  A compact orientable two-surface $\cal T$ is {\sl
strongly outer trapped} if on the two-surface, $\theta < 0$ for an
outer-directed hypersurface-orthogonal null congruence.  The next lemma
is valid even with vanishing expansion but the proof is simpler if we
assume that the expansion is strictly positive.

\noindent{\bf Lemma 2}. Let  $({\cal M}, {\rm g}_{ab})$ be an
asymptotically flat spacetime that satisfies ANEC and has a simply
connected Cauchy surface $\Sigma$. Then no surface ${\cal T}$, strongly
outer trapped with respect to ${\scri }_{\alpha }$, intersects
${J}^-({\scri}_{\alpha }^+)$.

{\sl Proof of Lemma 2}. The proof uses essentially the same techniques
as of Props. 9.2.1 and 9.2.8 in Hawking and Ellis\cite{HawkingEllis}.
If ${\cal T}$ intersects ${J}^-(\scri_{\alpha}^+)$, there is a causal
curve connecting $\cal T$ to ${\scri}_{\alpha }^+$. Hence ${J}^+({\cal
T})$ intersects ${\scri}_{\alpha }^+$. By definition, ${\scri}_{\alpha
}^+$ is closed, and since the spacetime is globally hyperbolic,
${J}^+({\cal T})$ is closed.  Hence ${\scri}_{\alpha }^+ \cap
{J}^+({\cal T})$ is closed. If ${({\scri}_{\alpha }^+ \cap {J}^+({\cal
T}))}^{\bullet }$ is empty, then  ${\scri}_{\alpha }^+ \cap {J}^+({\cal
T})$ is also open as a subset of  ${\scri}_{\alpha }^+$, because
${\scri}_{\alpha }^+ \cap {J}^+({\cal T}) = \hbox{\rm
int}({\scri}_{\alpha }^+ \cap {J}^+({\cal T})) \cup {({\scri}_{\alpha
}^+ \cap {J}^+({\cal T}))}^{\bullet}$.  But if ${\scri}_{\alpha }^+
\cap {J}^+({\cal T})$ is both closed and open, then ${\scri}_{\alpha
}^+$ is disconnected. However, by definition ${\scri}_{\alpha }^+$ is
connected. Therefore ${({\scri}_{\alpha }^+ \cap {J}^+({\cal
T}))}^{\bullet }$ is not empty and it follows that ${\scri}_{\alpha }^+
\cap {\dot J}^+({\cal T})$ is also nonempty; ${\dot J}^+({\cal T})$
intersects ${\scri}_{\alpha }^+$. A past directed null generator of
${\dot J}^+({\cal T})$ with future endpoint at a point of
${\scri}_{\alpha }^+$ must have past endpoint at ${\cal T}$ and can
contain no conjugate point. By {\sl Lemma 1}, the geodesic must be
outer directed from ${\cal T}$. But ANEC and $\theta < 0$ imply that
every outward null geodesic from the outer trapped surface ${\cal T}$
has a conjugate point within finite affine parameter length.  This is a
contradiction, because the generators of ${\dot J}^+({\cal T})$ have
infinite affine parameter length. Q.E.D.

{\sl Proof of Theorem 1}. Consider the universal covering space
\cite{cover} $\pi:{\cal M}\rightarrow M$ and the corresponding
spacetime $({\cal M},{\rm g}_{ab})$, with ${\rm g}_{ab}$ the pullback
of $g_{ab}$ to ${\cal M}$ by $\pi$.  By construction ${\cal M}$ is
simply connected, and any point in $M$ has a simply connected
neighborhood $A$ whose inverse image $\pi^{-1}(A)$ is the disjoint
union of open simply connected sets in ${\cal M}$.  Each of these
copies of $A$ in ${\cal M}$ corresponds to a homotopically distinct way
of reaching $A$ from a fiducial point of $M$, and we can choose the
fiducial point to lie on ${\scri}^+$. The projection $\pi$, restricted
to any single copy of $A$, is an isometry.

Since the open neighborhood $U$ of ${\scri}(M)$ is chosen to be simply
connected and $M$ itself is not simply connected, $U$ will be covered
by multiple copies of itself in ${\cal M}$, which will therefore have
multiple asymptotic regions. Let ${\cal U}_0\subset {\cal M}$ be one of
these copies, an open connected neighborhood of a single asymptotic
region of ${\cal M}$. Construct a partial conformal completion
$(\widetilde{\cal M}_0, \tilde{\rm g}_{ab})$ by adjoining a single copy
of ${\scri}(M)$ to ${\cal U}_0$.  Then $({\cal M}, {\rm g}_{ab})$, with
one asymptotic region singled out, satisfies the requirements of the
Lemmas.

Suppose the theorem is false. Then there is a causal curve $\gamma$ in
$M$, from ${\scri}^-(M)$ to ${\scri}^+(M)$, which is not deformable to
$\gamma _0$ relative to ${\scri}(M)$. The curves $\gamma_0$ and
$\gamma$ can be lifted to curves $\Gamma_0$ and $\Gamma$ in ${\cal M}$
that meet the same point of ${\scri}_0^+$.  Because the construction of
${\cal M}$ assigns distinct points to homotopically different ways of
reaching the same point of $M$, the curves $\Gamma_0$ and $\Gamma$ will
join ${\scri }_0^+({\cal M})$ to different copies of the asymptotic
region ${\scri}^-(M)$.  Because $\gamma_0$ lies in the simply connected
neighborhood $U$ of ${\scri}(M)$, $\Gamma_0$ will lie in the
neighborhood ${\cal U}_0$ of ${\scri}_0({\cal M})$, while $\Gamma $
will join ${\scri}_0^+({\cal M})$ to another copy of $U$.  In this
second asymptotic region, it will intersect spheres of arbitrarily
large radii.

These large spheres appear outer
trapped as seen from the first asymptotic region, ${\cal U}_0$: Let
$\overline \Sigma$ be the covering space of a
Cauchy surface $\Sigma$ of $M$ and let ${\cal S}$ be a sphere in an
asymptotic region of $\overline \Sigma$ different from the one
containing ${\cal U}_0$.  If we define outer-directed curves from any
sphere ${\cal S}'$ to be those that reach ${\cal U}_0$ without
intersecting ${\cal S'}$ a second time, then the outer directed curves
from ${\cal S'}$ are curves from its concave surface --- curves that
would ordinarily be called inner directed by an observer in the
asymptotic region near ${\cal S'}$.  Since the spacetime is
asymptotically flat, one can always pick ${\cal S'}$ so that the outer
directed null congruence has $\theta <0$.  As $\Gamma$ is causal, this
implies that there are strongly outer trapped surfaces that intersect
$J^-({\scri}_0^+({\cal M}))$. But this contradicts Lemma 2.  Hence any
causal curve $\gamma $ from ${\scri}^-$ to ${\scri}^+$ must be
deformable to $\gamma_0$. Q.E.D.

An alternate proof to the more standard proof given above can be obtained
using recent techniques
developed by Penrose, Sorkin and Woolgar \cite{psw} to prove a
spacetime version of the positive energy theorem.
They construct a partial ordering of all causal curves joining a given
generator of ${\scri}^-$ to a given generator of ${\scri}^+$, calling a
curve {\sl faster} if it arrives earlier at ${\scri }^+$ and leaves
later from ${\scri }^-$.  A fastest curve is a null geodesic without
conjugate points if the curve does not lie on ${\scri}$.
The same argument, restricted to curves in a
given homotopy class shows the existence of a null geodesic without
conjugate points lying in that homotopy class and joining ${\scri}^-$
to ${\scri}^+$ as needed to prove Theorem 1.

The consequences of Theorem 1 can be seen by considering a globally
hyperbolic spacetime with noneuclidean topology, assumed for simplicity
to have one asymptotic region. Its universal cover will be a spacetime
with multiple asymptotic regions. Suppose that an observer wishes to
probe the topology of her spacetime and communicate the results of her
measurements to a distant observer near ${\scri}^+$. In order to detect
a topological geon, her path or the path of her communication must
traverse  the geon and exit to ${\scri}^+$; but this is forbidden by
the theorem.  Only observers and light rays that do not loop around a
factor of the topology can communicate with ${\scri}^+$, and such
causal curves do not detect the existence of noneuclidean topology.
Thus general relativity prevents one from {\it actively} probing the
topology of spacetime.

However, note that one can {\it passively} observe that topology by
detecting light that originates at a past singularity.  This is in
keeping with the cosmic censorship conjecture which allows an observer
to see a singularity in her past; similarly, the active topological
censorship theorem proved above allows light rays to pass through a
point $x$ and then traverse homotopically distinct paths to a distant
observer if the rays originate at a past singularity.  The $RP^3$ geon
of the identified Schwarzschild geometry provides an example. In
Fig.~\ref{pendiag}, an observer $O$ outside the black hole can
passively detect the topology, receiving signals that traverse the
homotopically distinct paths $c$ and $c'$ from a point $x$ of the
nonsimply connected Cauchy surface $\Sigma$.  Followed back to the
past, these null geodesics eventually hit the singularity at $r=0$. In
accordance with the theorem, to passively detect the topology one must
see a signal that originates in a white hole rather than $\scri ^-$
(and hence at a singularity if cosmic censorship holds and if singularities
are generically spacelike).

In an earlier version of this paper\cite{fsw} we reported a result
which we ascribed to Schoen and Yau, stating that passive detection of
spacetime topology is allowed only for a restricted set of topologies:
all nontrivial topology due to a $K(\pi,1)$ factor is passively
censored. As Gregory Burnett \cite{burnett} has shown, this is false: there are
spacetimes in which $K(\pi,1)$ factors are passively observable.
Theorem 2 of Ref.\cite{fsw} stated:

\noindent{\bf (Theorem 2)}. Given any asymptotically flat initial data
set $(\Sigma, h_{ab}, p_{ab})$ with sources which obey the dominant
energy condition, all nontrivial topology due to a $K(\pi,1)$ prime
factor is surrounded by a two-sphere which is an apparent horizon.

The conclusion that the topology was unobservable arose from our
misinterpretation of ``apparent horizon''. It is standard in the
relativity literature (e.g. Hawking and Ellis\cite{HawkingEllis}
and Wald\cite{Wald}) to use this term
as shorthand for ``future apparent horizon''. In the above
theorem, it refers to {\it either} a future {\it or} a past apparent
horizon.
Therefore, one can only conclude that the $K(\pi,1)$ factors are either
within black holes or white holes. This conclusion already follows from
the active topological censorship result.

Finally, note that the $RP^3$ geon is a counterexample to any hope
that passive topological censorship holds in general. Combining this
result with Burnett's example one is led to the conjecture that
{\it all topologies are passively observable}.

To prove active topological censorship, we used the averaged null
energy condition, the weakest of the standard conditions.  However, it
is clear that any energy condition which implies that outward directed
null congruences with $\theta<0$ have conjugate points will suffice in
our proof. Several such weaker conditions are known\cite{borde}. This
suggests that the theorem may hold for semiclassical gravity, in which
the source is the expectation value of a renormalized stress tensor.
There are now several results which rely on conditions weaker than the
averaged weak or averaged null energy conditions and which may be valid
in a semiclassical theory:   the chronology protection theorem of
Hawking \cite{hawk}, the Penrose-Sorkin-Woolgar proof of a positive
energy theorem \cite{psw}, Gannon's theorem, and our active topological
censorship theorem.  Wald and Yurtsever  \cite{wy}  show that ANEC is
violated by the renormalized stress tensor of free fields in generic
spacetimes, and the question is whether there is a condition weak
enough to be satisfied by the semiclassical stress tensor and strong
enough to enforce the theorems.

\acknowledgments The authors are indebted to R. Geroch, R. Sorkin and
T. Jacobson for comments and discussions.  The work was supported by
NSERC and by the National Science Foundation under Grants PHY91-0593
and PHY89-04035.

\begin{references}
\bibitem{strongE} The weak energy condition is $T_{ab}w^aw^b\geq 0$ for
all timelike vectors $w^a$. The dominant energy condition is the weak
condition with the additional requirement that $T^{ab}w_b$ be causal.
The strong energy condition is $(T_{ab}-{1\over 2 }Tg_{ab}) w^a w^b
\geq 0$ for timelike vectors $w^a$.
\bibitem{cover} The universal covering manifold ${\cal M}$ of a smooth
path-connected manifold $M$ can be constructed as follows.  Pick a
point $x_0\in M$ and consider the set of smooth paths
$P=\{c:[0,1]\rightarrow M|c(0)=x_0\}$.  A projection map
$\pi:P\rightarrow M$ is defined by  $\pi(c(t))=c(1)$.  Let ${\cal M}$
be $P$ modulo the equivalence relation, $c_1\sim c_2$ if and only if
$c_1(1)=c_2(1)$ and $c_1$ is homotopic to $c_2$ with endpoints fixed.
The projection map $\pi$ is then well defined and smooth as a map
$\pi:  {\cal M}\rightarrow M$.
\end{references}
\end{document}